\documentclass[12pt,preprint]{aastex}

\shorttitle{The Cosmic Infrared Background at 1.25 and 2.2~$\mu$m}
\shortauthors{Cambr\'esy et al.} 

\begin{document}

\title{The Cosmic Infrared Background at 1.25~$\mu$m and
	2.2~$\mu$m using DIRBE and 2MASS: a contribution not due to galaxies ?}

\author{L. Cambr\'esy and W.T. Reach}
\affil{Infrared Processing and Analysis Center, California Institute of
Technology,\\ M/S 100-22, Pasadena, CA 91125}
\email{laurent@ipac.caltech.edu, reach@ipac.caltech.edu}
\author{C.A. Beichman}
\affil{Jet Propulsion Laboratory, California Institute of Technology,\\
M/S 180-703, Pasadena, CA 91109}
\email{chas@pop.jpl.nasa.gov}
\author{T.H. Jarrett}
\affil{Infrared Processing and Analysis Center, California Institute of
Technology,\\ M/S 100-22, Pasadena, CA 91125}
\email{jarrett@ipac.caltech.edu}

\begin{abstract}
Using the 2MASS $2^{\rm nd}$ Incremental Data Release and the
Zodiacal-Subtracted Mission Average maps of COBE/DIRBE, we estimate
the cosmic background in the $J$ (1.25~$\mu$m) and $K$ (2.2~$\mu$m) bands using
selected areas representing $\sim 550$~deg$^{2}$ of sky.
We find a $J$ background of $22.9\pm7.0$~kJy~sr$^{-1}$
($54.0\pm16.8$~nW~m$^{-2}$~sr$^{-1}$) and a $K$ background of
$20.4\pm4.9$~kJy~sr$^{-1}$ ($27.8\pm6.7$~nW~m$^{-2}$~sr$^{-1}$).
This large scale study shows that the main uncertainty comes from
the residual zodiacal emission. 
The cosmic background we obtain is significantly higher than integrated
galaxy counts ($3.6\pm0.8$~kJy~sr$^{-1}$ and $5.3\pm1.2$~kJy~sr$^{-1}$
for $J$ and $K$, respectively), suggesting either an increase of the galaxy
luminosity function for magnitudes fainter than 30 or the existence of another
contribution to the cosmic background from primeval stars, black holes, or
relic particle decay.
\end{abstract}


\section{Introduction}
The cosmic infrared background (CIB) is an important cosmological constraint
on the star formation history in the Universe \citep{DAH+98,GLP00}. 
The first detection of the CIB was reported by \citet{PAB+96}
for sub-millimetric wavelengths using data from the Far Infrared Absolute
Spectrometer (FIRAS) on-board the Cosmic Background Explorer (COBE).
Shorter wavelengths have been investigated with the help of the COBE Diffuse
Infrared Background Experiment (DIRBE) from 240~$\mu$m to 1.25~$\mu$m.
\citet{HAK+98} measured the CIB at 240 and 140~$\mu$m and set upper
limit for other wavelengths.
\citet{LHRT00} proposed an estimation of the CIB at 100~$\mu$m. 

The CIB is the signal that remains after subtracting from the total celestial
brightness the emission from the interstellar medium, stars and interplanetary
dust (scattering and emission). However, estimating the contamination due to
these three foreground components is challenging. The interstellar medium is
the main contaminant for long wavelengths whereas starlight and zodiacal
light dominate at wavelengths studied in the present paper ($J=1.25$~$\mu$m
and $K=2.2$~$\mu$m).

Stellar population models were used to estimate the brightness of starlight
in the large beam DIRBE data \citep{AOW+98} but this technique led only to
upper limits. Detections of the CIB at 3.5~$\mu$m ($L$) and 2.2~$\mu$m
(K) were reported by \citet{GWC00} using near-infrared observations of a
$2^\circ \times 2^\circ$ dark spot to measure the brightness of stars in a
few DIRBE beams. \citet{Wri01} confirmed the result at 2.2~$\mu$m and proposed
a weak limit for 1.25~$\mu$m ($J$) using data from the 2 Micron All-Sky survey
(2MASS) for 4 dark spots.
\citet{MCF+00} estimated the CIB from 1.4~$\mu$m to 4~$\mu$m with the
Near Infrared Spectrometer (NIRS) on-board the InfraRed Telescope in Space
(IRTS). Although NIRS provided a point source catalog limited to stars
brighter than 7.5~mag at 2.24~$\mu$m, significant stellar
contamination remains and the CIB estimate is strongly dependent on the 
stellar population model used to remove contribution from fainter stars.

This work presents a large scale study of the CIB using 2MASS data for
1400~deg$^{2}$ of the sky in order to estimate accurately the stellar
contribution to the surface brightness observed by DIRBE. Section~\ref{data}
describes the 2MASS and DIRBE data. The method used to compare these two data
sets is presented is Section~\ref{method}, and the following section presents
the determination of the CIB for $J$ and $K$ with the associated uncertainties.
Section~\ref{discussion} is dedicated to the comparison with previous results
and with galaxy counts.

\section{Data}
\label{data}
\subsection{2MASS}
The 2MASS second Incremental Data Release from the 2MASS survey covers
48\% of the sky and contains $1.6\times 10^8$ stars \citep{CSV+00}. The Point
Source Catalog completeness limits are 15.8 and 14.3~mag with a
signal-to-noise ratio greater than 10:1 for $J$ and $K_s$, respectively.
No photometry is provided for stars brighter than 4-5~mag
(see Sect.~\ref{verybright}).

Since the catalog of the whole release is $\sim 7$ GB in binary format,
we choose to work on integrated maps in which each pixel corresponds to the
integrated flux in a $5' \times 5'$ box. These maps are constructed from the
point source catalog clipped to remove sources fainter than the completeness
limit. Magnitudes are converted to flux density using the flux for zero
magnitude $F_0(J)=1603$~Jy defined by \citet{CRL85} and $F_0(K_s)=698$~Jy
which is an extrapolation of the $J$ value to the $K_s$ filter assuming a
black body at 9700~K (Vega photometric system). Uncertainties on these
photometric zero points are about 2\% and do not affect our results.
Each $5' \times 5'$ box contains the sum of the point source fluxes in the
area. 

The resulting image in an Aitoff (equal area) projection consists of
$4321 \times 2161$ pixels.
This size is reasonable and the image can be handled globally. Errors resulting
from this pixelization are negligible since $5'$ is small compared to
the DIRBE beam of $1^\circ$ (see below).

We should point out that the 2MASS $K_s$ filter (2.17~$\mu$m) and the DIRBE
$K$ filter (2.2~$\mu$m) are slightly different. \citet{PMK+98} discuss the
$K$ to $K_s$ transformation and present a list of photometric standards in
both colors. Moreover, as \citet{PM94} point out, the $K_s$ bandpass is
intermediate between $K$ and $K'$ and rough corrections of $K_s$ to $K$
can be obtained by averaging between $K'$ and $K$. This transformation is
well defined by \citet{WC92}: $K'-K = (0.22\pm0.03)(H-K)$  for $0<H-K<0.4$
($K'-K \sim 0.07$ otherwise). Hence, for a K5 spectral type star we would
obtain $K_s-K \sim 0.01$ for example. This correction is negligible compared
to uncertainties described further in the paper and differences between 2MASS
and DIRBE filters will be ignored in the following analysis.

\subsection{DIRBE}
Cryogenic DIRBE operation took place from 1989 December to 1990 September.
During these 10 months, the sky was observed at the rate of half of the sky
per week in 10 bands from 1.25~$\mu$m to 240~$\mu$m \citep{HAK+98}.
The DIRBE instrument was designed to make accurate absolute sky brightness
measurements, with a straylight rejection of $<1$~nW~m$^{-2}$~sr$^{-1}$ and
an absolute brightness calibration uncertainty of 0.05 and
0.03~nW~m$^{-2}$~sr$^{-1}$, at 1.25 and 2.2~$\mu$m, respectively.

The zodiacal light is the first component along the light of sight and must
be removed using a dust interplanetary model \citep[e.g.][]{KWF+98}. However
these models are not unique (see \citealp{Wri98,GWC00} for another model). 
Artifacts still remain.
In the following analysis, we use the Zodi-Subtracted Mission Average (ZSMA)
maps produced by the DIRBE team \citep{HKW98}.
The zodiacal light intensities were subtracted week by week and the residual
intensity values were averaged to create the ZSMA maps.

\begin{figure}
\plottwo{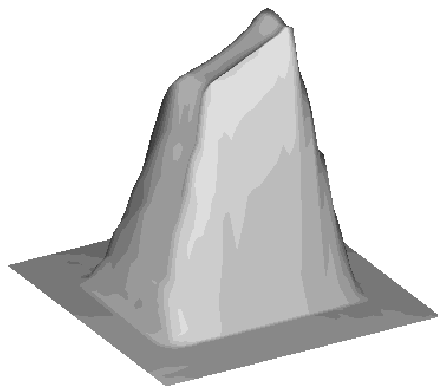}{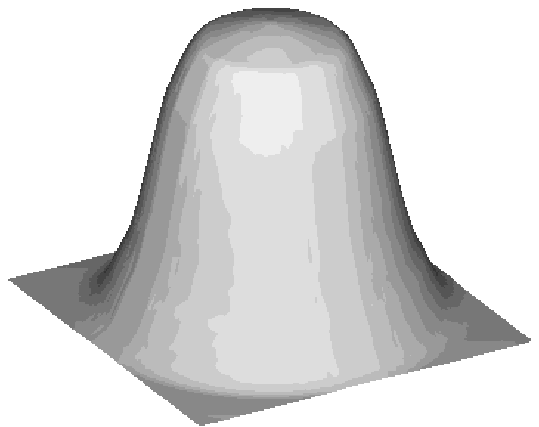}
\caption{{\em Left}: effective DIRBE beam for daily maps (FWHM$=42'$)
	at 1.25~$\mu$m. {\em Right}: averaged beam for annual map (FWHM$=1^\circ$).}
\label{dirbeJbeam}
\end{figure}

To compare DIRBE and 2MASS, we need a precise knowledge of the DIRBE
beam in both $J$ and $K$ bands. 
The $42'$ beam size commonly mentioned in the literature is only valid for 
daily maps, not the annual maps used here. The beam for the $J$ band is
presented in the Figure~\ref{dirbeJbeam}. This is an effective beam profile
(provided with DIRBE maps),
which measures the relative response of DIRBE to a point source, including
the effects of sky-scanning and data-sampling rates.
The beam for the annual maps corresponds to an average of the effective beam
for all orientations (Fig.~\ref{dirbeJbeam}). Similar results are obtained
for $J$ and $K$ and the full width half maximum of both beams is
$\sim 1^\circ$ (solid angle $\sim 0.78$~deg$^{2}$).

\section{Method}
\label{method}
\subsection{2MASS integration on DIRBE pixels}
First, we exclude from this study regions of the sky for which we know that 
the CIB cannot be straightforwardly extracted, for example due to high
stellar density fields, interstellar cirrus (which is responsible for
emission and scattering), and regions of residual zodiacal light structures
in the DIRBE maps.

We keep only regions that satisfy the following criteria:
\begin{itemize}
\item high galactic latitude: $|b| > 40^\circ$
\item low DIRBE 240~$\mu$m flux in order to eliminate diffuse emission or
	scattered light from cirrus clouds: $I_{240} < 3$~MJy~sr$^{-1}$
\item high ecliptic latitude: $|\beta| > 30^\circ$
\item exclusion of the Magellanic clouds
\end{itemize}

\begin{figure}
\plotone{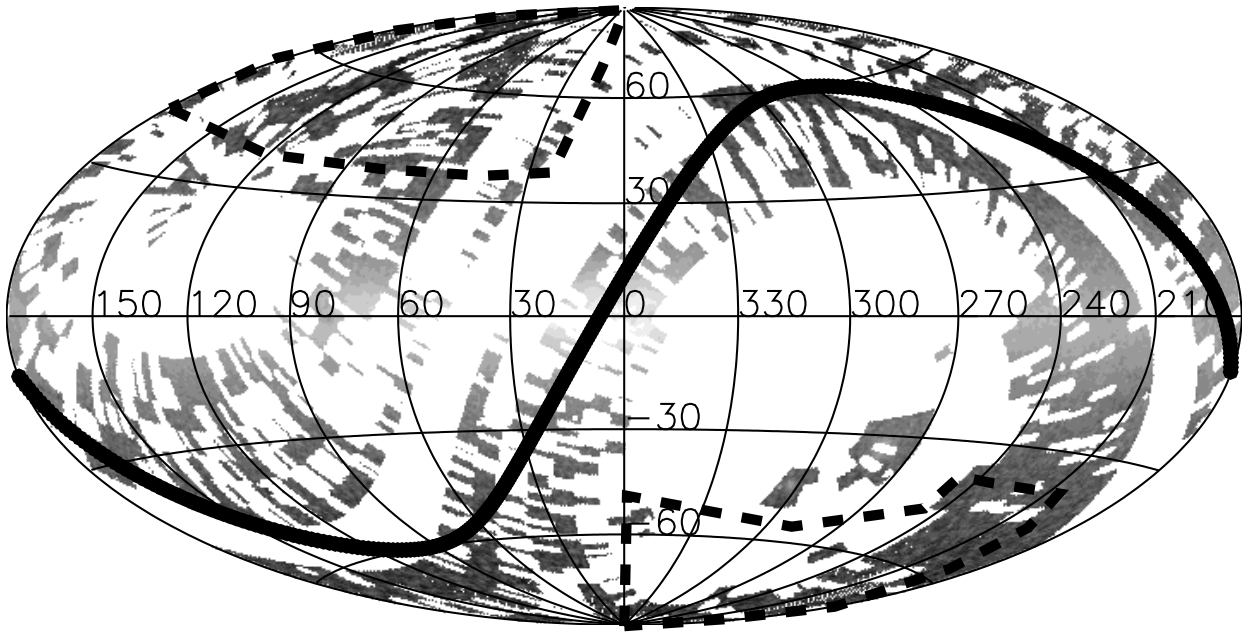}
\caption{Aitoff projection in galactic coordinates of the integrated 2MASS map
	($J$) on DIRBE pixels. The large line represents the ecliptic plane.
	This study is limited to the areas inside the dashed curves (i.e. high
	galactic and ecliptic latitudes) and non-zero pixels correspond to
	$\sim 1400$~deg$^{2}$.
}
\label{integJ}
\end{figure}

The most straightforward way to remove the stellar component (2MASS) from the
DIRBE flux is to work directly in the COBE coordinate system (projection on
a cube) and to integrate the 2MASS maps on each DIRBE pixel. However, since
the 2MASS data still contain coverage holes, we kept only those that are
completely covered by 2MASS. The effective useful region covers finally about
1400~deg$^{2}$ (Fig.~\ref{integJ}).

\subsection{Bright stars}
\label{verybright}
In the remaining 1400~deg$^{2}$ piece of sky, $\sim 25$\% of pixels are
contaminated by bright stars. Stars bright enough to be detected as point
sources in DIRBE are saturated in 2MASS. Consequently their photometry cannot
be derived from 2MASS to be subtracted from DIRBE.

Moreover, the DIRBE beam has a size of $1^\circ$, but the sampling of the data
is $22'$. Several pixels are therefore contaminated by a bright star, and this
contamination depends on the star position in the beam. To identify
pixels affected by these stars we use a median filter which deals with the 
2MASS holes. 
A pixel is considered as contaminated if it is brighter than
the threshold, $\mathcal{T}$, defined by the median plus 
$2\sigma$, $M + 2\sigma$, of the surrounding pixels in a $3^\circ$ radius
circle of the DIRBE$-$2MASS map:
\begin{equation}
\label{eq_med}
\mathcal{T}^i = M^i\left(\sum_{i,r=3^\circ}{\rm DIRBE}-{\rm 2MASS}\right) + 2\sigma^i\left(\sum_{i,r=3^\circ}{\rm DIRBE}-{\rm 2MASS}\right)
\end{equation}
Pixels fainter than $M-2\sigma$ are also removed to avoid bias.
In order to validate this operation, we use the 2MASS Bright Infrared Star 
Compilation (Tam et al. 1999\footnote{http://spider.ipac.caltech.edu/staff/raymond/2mass/birsc/birsc.html}) which contains photometry for stars as
bright as $-4$ mag in $JHK_s$. This compilation consists of data taken from
the literature and photometry extrapolations from IRAS, MSX or 2MASS.
Among the 343 stars brighter than $J=5$ mag in our regions of interest, 11
are not identified with the median filter (Eq.\ref{eq_med}). However, these
remaining contaminated pixels are rejected by the robust linear fitting of
DIRBE/2MASS brightness (see Fig.~\ref{pointJK}).

After removing bright stars, 1040~deg$^{2}$ of the sky remain.

\subsection{Faint stars}
The 2MASS integrated brightness maps were limited to the formal completeness
limits of 2MASS ($J^{\rm lim}=15.8$~mag, $K_s^{\rm lim}=14.3$~mag). A model is
required to estimate the contribution of the fainter stellar population with
respect to the position in the sky. We use a model in which the number and
distribution of Milky Way stars as seen in the near to mid-infrared is
adapted from the \citet{BS80a} optical star count model. We have employed the
discrete population formalism of \citet{Eli78}, \citet{JAHR81}, \citet{GJ87}
and \citet{Jar92}.
The model includes the class III (evolved giant), class IV (subdwarfs) and
class V (main sequence) stellar populations. These are further divided into
disk and spheroid spatial distributions (ala Bahcall \& Soneira). The stars
are discretely
binned according to their spectral types, ranging from the hottest O stars to
the coolest M dwarfs, giving a total of 22 separate spectral bins for the main
sequence, and 12 for the luminous giants (ranging from G2 to M7 giant). The 
optical/infrared colors, fluxes and luminosity function per spectral type are
based on empirical data (e.g., Koornneef 1983; 2MASS, ISO and IRAS).
Interstellar extinction is applied as a smooth exponential function of
Galactic position, characterized by a scale height and disk length. The model
parameters were tuned using deep optical and infrared star counts.
The resultant model was validated using the 2MASS survey down to $J = 15$ 
and $K_s = 14$~mag ($\sim 2$~mJy).
\begin{figure}
\epsscale{0.5}
\plotone{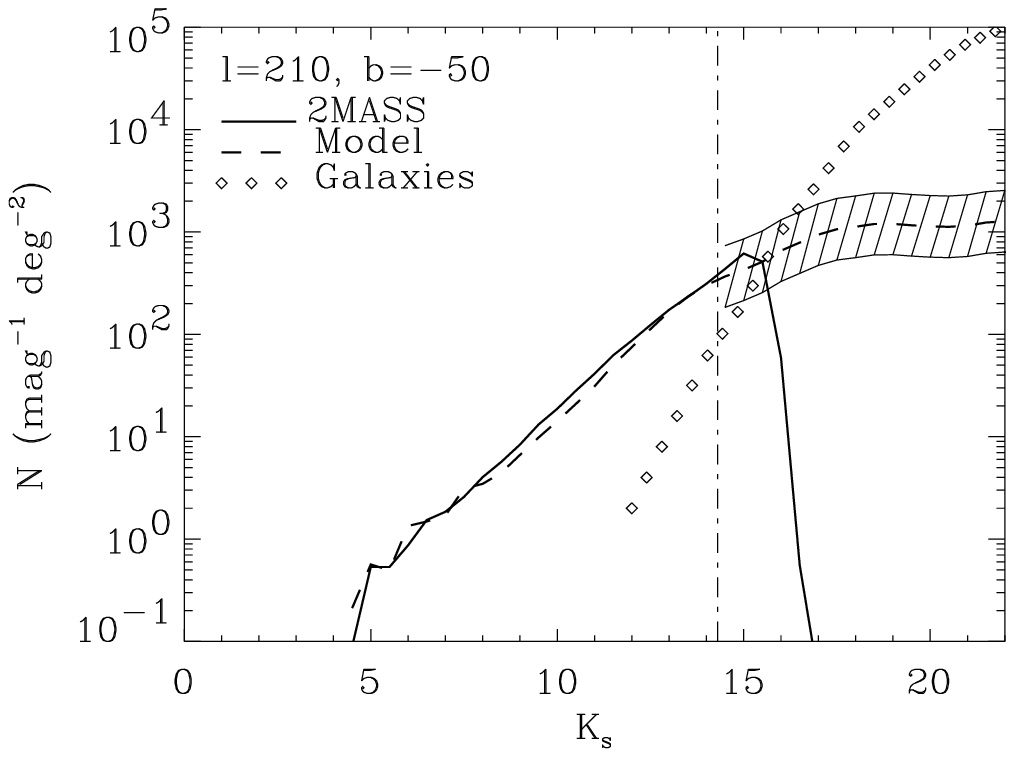}
\caption{$K_s$ source counts from 2MASS (solid line), from the model used for
	faint stars (dashed line) and galaxy counts (diamond) from
	\citet{GSFC97}. The shaded area corresponds to the model uncertainties.
	The vertical line shows the cutoff used in the 2MASS catalog, the
	model provides us counts for fainter magnitudes. Note that at this
	location ($l=210^\circ$, $b=-50^\circ$), galaxies dominate the counts
	for $K_s > 16$.}
\label{cmodel}
\end{figure}

The result of the model for a typical high galactic latitude field
($l=210^\circ$, $b=-50^\circ$) is presented in Figure~\ref{cmodel}. As expected,
2MASS counts are complete at least to $K_s=14.3$~mag. According to the model,
less than 3\% (0.1\%) of the total stellar energy is contained in stars
fainter than $K_s = 14.3$ (19). The model is useful for magnitudes ranging
from 14.3 to $\sim 20$ in $K_s$. Assuming a conservative number density
uncertainty of a factor of 2 for these faint magnitudes, the resulting error
will not exceed 3\% of the total stellar flux.

\section{Result}
\label{result}
\subsection{DIRBE / 2MASS correlation}
\begin{figure}
\epsscale{1.0}
\plottwo{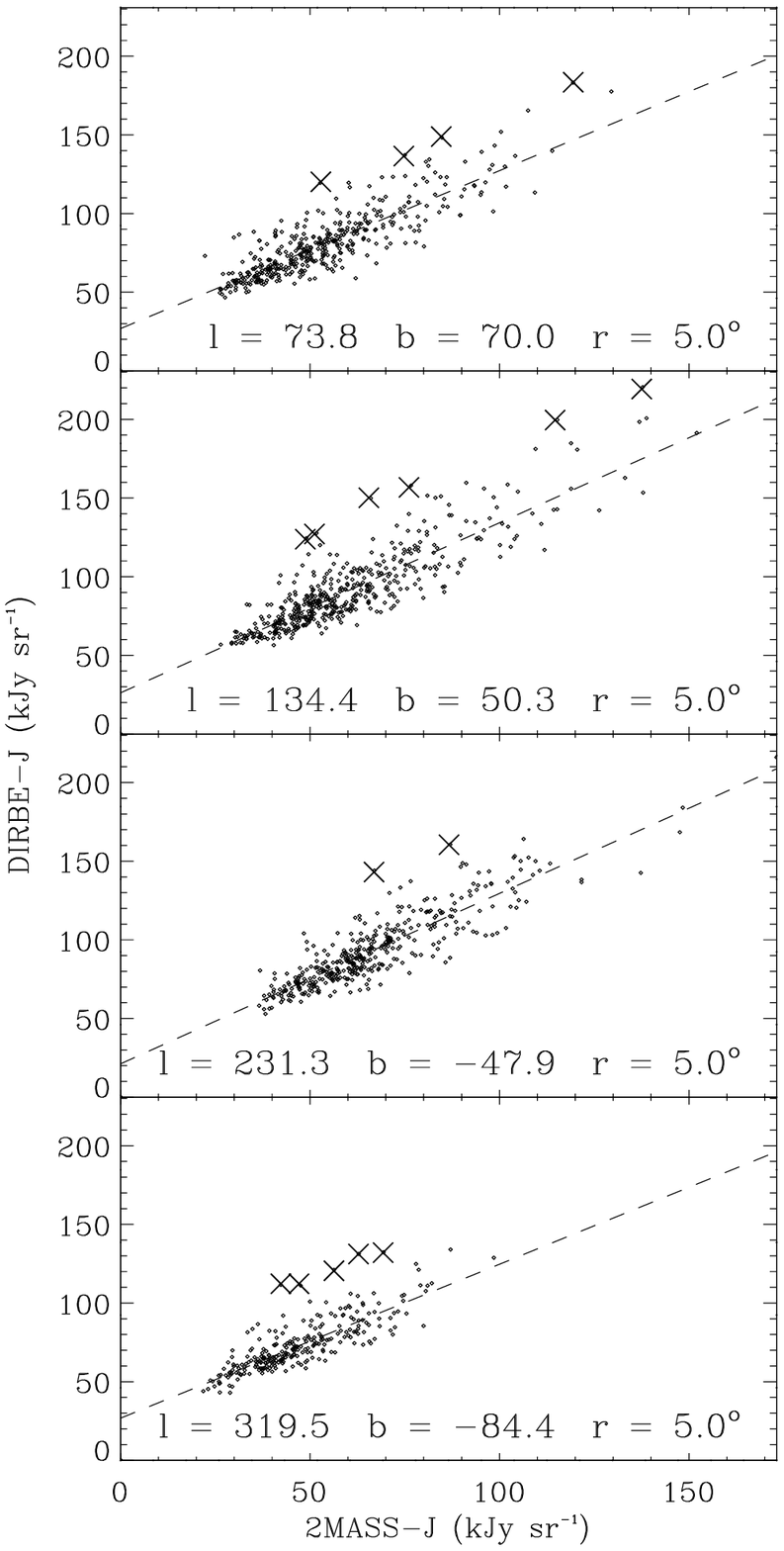}{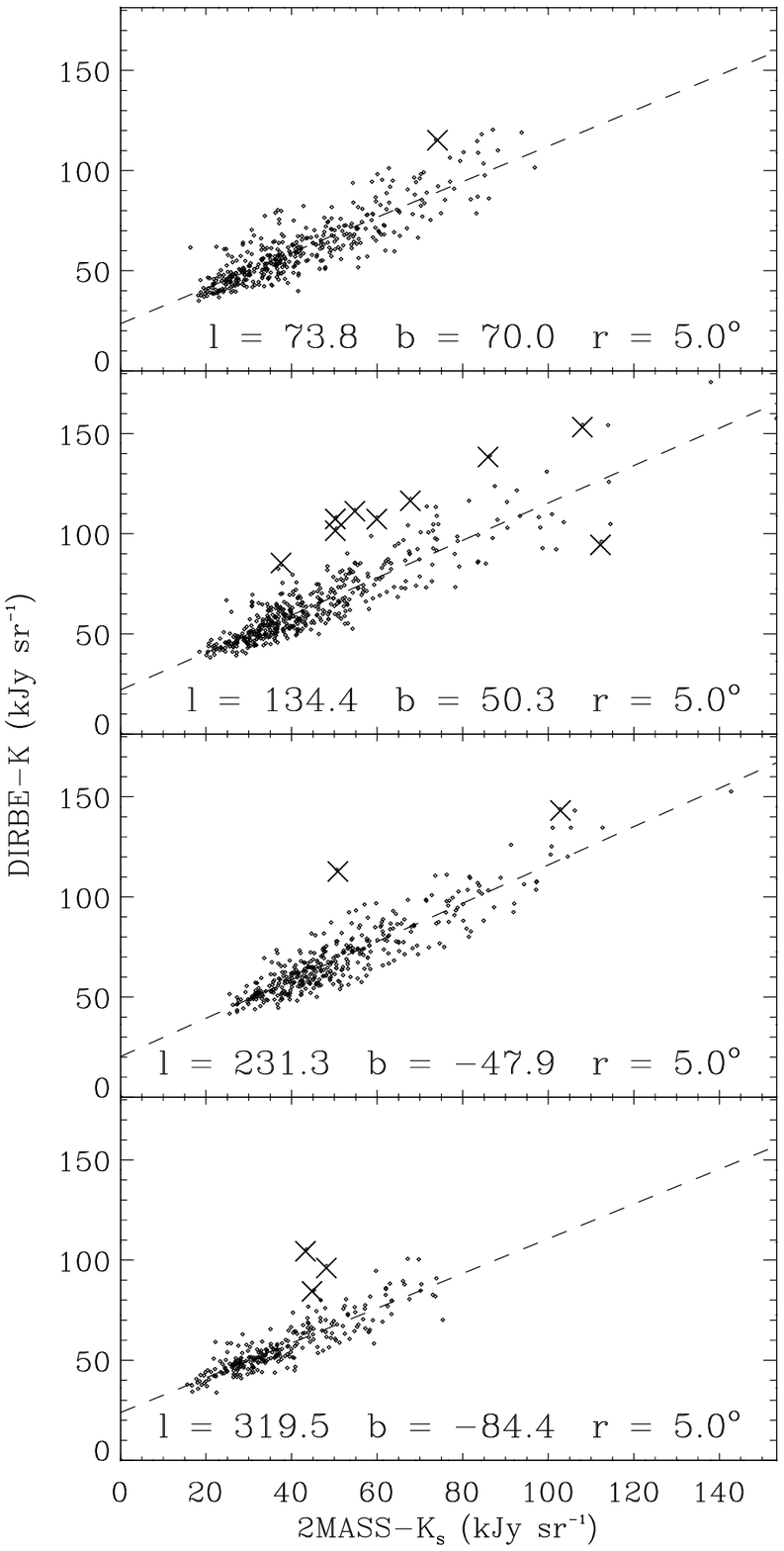}
\caption{DIRBE versus 2MASS for 4 regions of $5^\circ$ radius.
Pixels contaminated by bright stars have been removed. Crosses correspond to
pixels rejected by the $3\sigma$ robust fitting. See Table~\ref{tab1}
for main characteristics.}
\label{pointJK}
\end{figure}

We choose 4 areas with a radius of $5^\circ$ in order to illustrate the
correlation between DIRBE and 2MASS. Pixels contaminated by bright stars have
been filtered. Figure~\ref{pointJK} show the correlations for $J$ and $K$ bands.
The main characteristics for each region are summarized in Table~\ref{tab1}.
We note that slopes are not exactly 1 ($\sim 1.07$ for $J$, $\sim 0.94$ for
$K$). This arises from the different calibration strategy used by DIRBE and
2MASS: DIRBE is calibrated with Sirius, 2MASS with a list of calibration
stars of many different spectral types \citep{CSV+00}. These calibration
differences affect only the slope of the DIRBE/2MASS relation, not the value
of the intercept which is the measure of the CIB.

\begin{deluxetable}{ccccccccc}
\tablecaption{DIRBE/2MASS correlations for the four areas\label{tab1}.}
\tablewidth{0pt}
\tablehead{
	\colhead{band} &
	\colhead{$l$} &
	\colhead{$b$} &
	\colhead{nb\tablenotemark{a}} &
	\colhead{slope} &
	\colhead{intercept\tablenotemark{b}} &
	\colhead{corr\tablenotemark{c}} &
	\colhead{faint stars\tablenotemark{d}} &
	\colhead{CIB\tablenotemark{b}} 
	}
\startdata
$J$ &  73.84 & 70.02 & 407 & $1.00\pm0.03$ & $26.9\pm1.6$& 0.87 & 1.44 & 25.5\\ 
$K$ &        &       & 409 & $0.89\pm0.02$ & $23.6\pm1.1$& 0.87 & 1.84 & 22.0\\
$J$ & 134.42 & 50.30 & 462 & $1.08\pm0.03$ & $26.2\pm1.7$& 0.88 & 1.68 & 24.4\\
$K$ &        &       & 472 & $0.93\pm0.02$ & $22.0\pm1.0$& 0.89 & 2.34 & 19.8\\
$J$ & 231.32 &-47.86 & 386 & $1.08\pm0.03$ & $21.2\pm1.9$& 0.88 & 1.81 & 19.2\\
$K$ &        &       & 389 & $0.96\pm0.02$ & $20.3\pm1.2$& 0.91 & 2.54 & 17.9\\
$J$ & 319.54 &-84.42 & 263 & $0.98\pm0.04$ & $26.8\pm1.8$& 0.82 & 1.64 & 25.2\\
$K$ &        &       & 260 & $0.87\pm0.03$ & $23.9\pm1.1$& 0.86 & 2.05 & 22.1\\
\enddata
\tablenotetext{a}{number of pixels}
\tablenotetext{b}{expressed in kJy~sr$^{-1}$}
\tablenotetext{c}{correlation coefficient}
\tablenotetext{d}{faint stars ($J>15.8$, $K_s>14.3$) contribution from the
					Jarrett model in kJy~sr$^{-1}$}
\end{deluxetable}
Correlation coefficients show that the correlation is good for both colors
(see Table~\ref{tab1}).
The CIB is obtained by removing the faint star contribution to the intercept
value.

\subsection{Zodiacal contamination}
\label{zodi}
Since the main error is suspected to come from the zodiacal
subtraction in the DIRBE maps, a representation of the CIB versus the ecliptic
latitude $\beta$ is useful.
For each pixel, we estimate the intercept of the DIRBE/2MASS linear
correlation in a $5^\circ$ radius area, and we remove the faint star model
from the DIRBE data. After this operation we select regions with the best
DIRBE/2MASS
correlation as defined in Table~\ref{tab_sel}. These values were obtained by
examination of the histogram of each quantity, and correspond to the peak
value divided by 4. This selection leads to $\sim 550$~deg$^{2}$ of
reliable regions in the sky.

\begin{figure}
\plotone{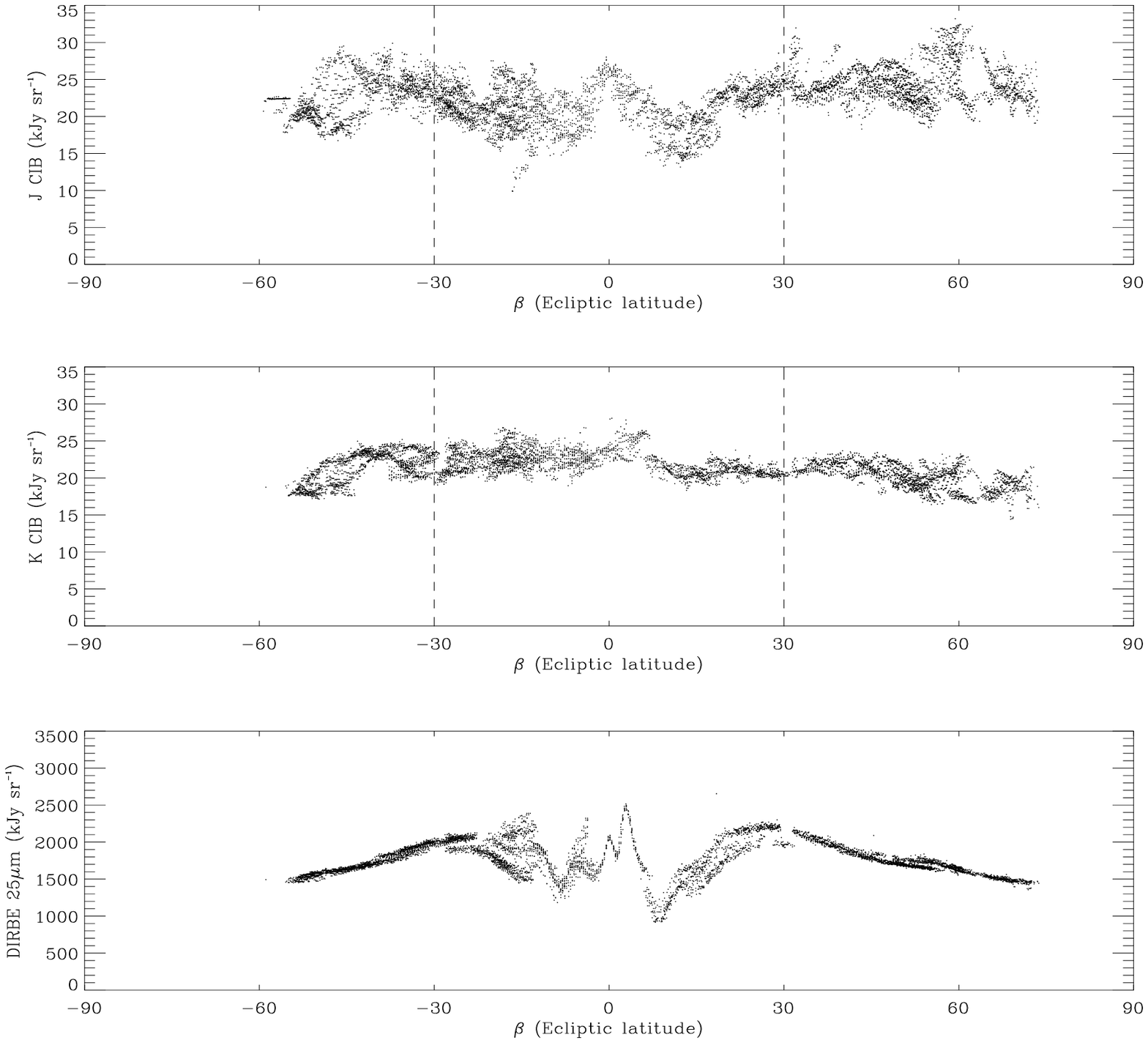}
\caption{CIB versus ecliptic latitude for $J$ (upper plot) and $K$ (middle
	plot). Only pixels with galactic $|b|>40^\circ$, good DIRBE/2MASS
	correlation and low DIRBE 240~$\mu$m are plotted. The lower plot
	represents the total DIRBE flux at 25~$\mu$m versus the ecliptic
	latitude for comparison. The same zodiacal light subtraction features
	appear in the 3 maps. The main uncertainty comes therefore from the
	zodiacal light and resulting CIB values are $22.9\pm7.0$~kJy~sr$^{-1}$
	($54.0\pm16.8$~nW~m$^{-2}$~sr$^{-1}$) and $20.4\pm4.9$~kJy~sr$^{-1}$
	($27.8\pm6.7$~nW~m$^{-2}$~sr$^{-1}$) for $J$ and $K$, respectively.}
\label{eclip}
\end{figure}

Figure~\ref{eclip} shows the resulting CIB versus ecliptic latitude for both
colors with an additional plot of the DIRBE 25~$\mu$m Zodi-subtracted surface
brightness. For this figure, we include both low and ecliptic latitude points.
The 25~$\mu$m band is the most sensitive to the zodiacal light and
contains subtraction residuals from the interplanetary dust model. 
The comparison of the 25~$\mu$m plot with the CIB plots confirms that the
scatter at low ecliptic latitude comes mainly from residual zodiacal effects.
\begin{deluxetable}{lll}
\tablecaption{Selection of the best DIRBE/2MASS correlation
	areas\label{tab_sel}.}
\tablewidth{0pt}
\tablehead{
	\colhead{Quantity} & \colhead{$J$} & \colhead{$K$}
	}
\startdata
number of pixels   & $\geq 80$         & $\geq 80$         \\
correlation coeff  & $\geq 0.72$       & $\geq 0.82$       \\
mean square dev.   & $\leq 200$        & $\leq 80$         \\
intercept $\sigma$ & $\leq 2.8$        & $\leq 1.6$        \\
slope $\sigma$     & $\leq 0.05$       & $\leq 0.035$      \\
slope              & $\in[0.97,1.125]$ & $\in[0.85,0.965]$ \\
\enddata
\end{deluxetable}

Systematic uncertainties associated with the zero point of the interplanetary
dust models are difficult to estimate. \citet{KWF+98} estimate them by choosing
the {\em largest} among three models at high ecliptic latitude.
Accordingly, a conservative estimate of the systematic uncertainties
in $J$ and $K$ is 6.25~kJy~sr$^{-1}$ and 4.4~kJy~sr$^{-1}$, respectively.
If the three models used to derive these numbers were equally distributed
in the space of possible models, the given uncertainties would correspond to
$1.6\sigma$. Unfortunately, the models are probably not equally probable and
it is hard to interpret these numbers in terms of a confidence level. To be
conservative, we assume they correspond to $1\sigma$. Theoretical
uncertainties in the zodiacal light model are further discussed by
\citet{DAH+98}.

Statistical uncertainties that comes from the DIRBE/2MASS correlation and from
the zodiacal light can be estimated directly from Figure~\ref{eclip} by
measuring the root mean square deviation of the distribution. The resulting
uncertainties are 2.7~kJy~sr$^{-1}$ for $J$ and 2.1~kJy~sr$^{-1}$ for $K$. A
summary of the different uncertainty contributions is presented in
Table~\ref{tab2}. The presence of fluctuations in the residual sky brightness,
especially in $J$, suggests a small contribution to the sky brightness by
starlight scattered off of interstellar dust grains.

\subsection{Local interstellar medium contamination}
Interstellar dust passing through the Solar System will scatter and reemit
absorbed sunlight. The volume within which sunlight is significant (compared
to the interstellar radiation field [ISRF]) is smaller than the size of the
Solar System itself: at visible wavelengths, sunlight exceeds the ISRF out to
the Oort cloud ($\sim 10^4$~AU), while at far-ultraviolet wavelengths,
sunlight exceeds the ISRF only out to 800~AU (at 0.2 $\mu$m) or less. 
The amount of interstellar dust close to the Sun was recently measured by the
Ulysses and Galileo spaceprobes \citep{GGM+94}; the volumetric cross-section
is $n_d\sigma_d\simeq 2.9\times 10^{-23}$~cm$^2$~cm$^{-3}$.
It was found that the size distribution is deficient in small particles
relative to the interstellar size distribution, such that the surface area
is dominated by particles with size $\simeq 0.4$ $\mu$m. 

To estimate the brightness of sunlight scattered by local interstellar dust,
we assume the particles detected by Ulysses are spread uniformly throughout
the Solar System. The emission from dust in the inner Solar System will have
a detectable dependence on solar elongation angle, and would have been
included in the DIRBE zodiacal emission model \citep{KWF+98}. Therefore,
we integrate from 3~AU outward in calculating the brightness of sunlight
scattered by interstellar dust and potentially contributing to an isotropic
background. The brightness is
\[ I_{\rm lism} =\int_3^\infty{ \omega \, n_d\sigma_d \, I_\odot \, \Phi \, dL}, \]
where $\omega$ is the albedo, $I_\odot$ is the solar intensity, and $\Phi$ is
the scattering phase function. Using the properties of a mixture of
astronomical silicates and graphite \citep{DL84}, the albedo in the 
$J$ and $K$ bands is 0.42 and 0.21, respectively. For the phase function, we
used a Henyey-Greenstein function with the asymmetry factor appropriate
for the same mixture of astronomical silicates and graphite ($g_J=0.15$,
$g_K=0.02$). The resulting brightness is
$I_{\rm lism} = 0.8$ and $0.3$~kJy~sr$^{-1}$ in the $J$ and $K$ bands,
respectively.

Uncertainty in the estimate of the brightness of the local interstellar dust
is due to three sources.
First, our lack of accurate knowledge of the composition of the interstellar
grains leads to a factor of approximately 2 uncertainty in their albedo; this
estimate is based on the difference between the albedos of silicate and
graphite grains \citep{DL84}.
Second, the lack of precise distinction between the emission that would be
effectively isotropic and that which would be incorporated into the zodiacal 
light model leads to a $\sim 20$~\% uncertainty; this is based on changing
the minimum integration distance from 3 to 2.5~AU.
Third, smaller particles that are not detected by Ulysses may be present
at larger distances from the Sun; these particles are likely to be
produce an insignificant scattering because of the weakness of sunlight
at large distances from the Sun.

To summarize, we estimate the CIB as follows:
${\rm CIB}(J) = 22.9 \pm 7.0$~kJy~sr$^{-1}$
				($54.0 \pm 16.8$~nW~m$^{-2}$~sr$^{-1}$) and 
${\rm CIB}(K) = 20.4 \pm 4.9$~kJy~sr$^{-1}$
				($27.8 \pm 6.7$~nW~m$^{-2}$~sr$^{-1}$).

\begin{deluxetable}{lcc}
\tablecaption{Uncertainties (values for $1\sigma$)\label{tab2}.}
\tablewidth{0pt}
\tablehead{
	\colhead{Quantity} &
	\colhead{$J$} &
	\colhead{$K$}
	}
\startdata
Bright Stars ($\leq5$ mag) & removed & removed \\
2MASS flux per DIRBE pixel \\ \ \ ($5<J\leq15.8$, $5<K_s\leq14.3$) &
				$4.2\times10^{-5}$ kJy~sr$^{-1}$ &
					$4.5\times10^{-5}$ kJy~sr$^{-1}$\\
Faint stars model \\ \ \ ($J>15.8$, $K_s>14.3$) & $<2$ kJy~sr$^{-1}$ &
							$<2$ kJy~sr$^{-1}$ \\
DIRBE flux & $2.2$ kJy~sr$^{-1}$ & $2.1$ kJy~sr$^{-1}$ \\
DIRBE/2MASS slope & $<10$\%, no consequence &
					$<10$\%, no consequence \\
$K \neq K_s$ & $\diagup$ & $2.5\times10^{-5}$ kJy~sr$^{-1}$\\
CIB scatter\\ \ \ (contains all stat. uncert.) & $2.7$ kJy~sr$^{-1}$ &
							$2.1$ kJy~sr$^{-1}$ \\
Zodiacal light model \\ \ \ (syst. uncert.) & $6.25$ kJy~sr$^{-1}$ &
							$4.4$ kJy~sr$^{-1}$ \\
Local Interstellar Medium & $1.6$ kJy~sr$^{-1}$ & $0.6$ kJy~sr$^{-1}$ \\ 
Total CIB uncertainty &  $7.0$ kJy~sr$^{-1}$ & $4.9$ kJy~sr$^{-1}$ \\
\enddata
\end{deluxetable}

\section{Discussion}
\label{discussion}
\subsection{Comparison with previous studies}
Previous estimations of the CIB in $K$ are $16.4 \pm 4.4$~kJy~sr$^{-1}$ and
$14.8 \pm 4.1$~kJy~sr$^{-1}$ by \citet{GWC00} and \citet{Wri01}, respectively. 
The difference between these values and the estimation given in this paper
are within the measurement uncertainties ($1\sigma$) and thus are in good
agreement. Most of the difference can be attributed to the zodiacal light
subtraction because we used the \citet{KWF+98} model and they used the
\citet{Wri98} model.

The estimation of the $J$ CIB can be compared to the weak limit of 
$12.0 \pm 6.3$~kJy~sr$^{-1}$ proposed by \citet{Wri01}. The discrepancy
with our value ($22.9 \pm 7.0$~kJy~sr$^{-1}$) can be explained by the zodiacal
light subtraction, since Wright would find 25.8~kJy~sr$^{-1}$ with the
\citet{KWF+98} dust interplanetary model.

\citet{KO00} have investigated the CIB fluctuations in the DIRBE maps from
1.25~$\mu$m to 4.9~$\mu$m. Fluctuations on the DIRBE beam scale are expected
to be 10-20\% of the total CIB flux (Kashlinsky, private communication). Their
values of $6.5^{+1.5}_{-2.9}$ and $4.3^{+1.2}_{-2.7}$~kJy~sr$^{-1}$ (92\%
confidence level) for $J$ and $K$ respectively are therefore in agreement
with our results in both bands.

\subsection{Galaxy counts}
Galaxy counts and the CIB are closely related: at least part of the CIB is 
due to the integrated surface brightness of galaxies for all redshifts. In the
past decade, many efforts have been made to obtain galaxy counts in a wide
range of magnitudes.
\citet{Jar01} performed large scale studies with 2MASS (2350~deg$^{2}$)
which provide galaxy counts for $8 \leq J \leq 15$ and $8\leq K_s \leq 14$. 
\citet{BLK98} obtained ground based observations up to $24^{\rm th}$
magnitude in $J$ and $K$ with the Keck 10m telescope. \citet{SDM+99} used
ESO~NTT telescope and obtained galaxy counts for $17 \leq J \leq 24$ and
$16 \leq K \leq 22$.
Deeper observations are possible with the Hubble Deep Field (HDF) of the Hubble
Space Telescope (HST) and \citet{TSW+99} reached the $30^{\rm th}$ magnitude
in the HST filters F160W and F110W which correspond roughly to $I+J$ and $H$,
respectively.

\begin{figure}[htb]
\plottwo{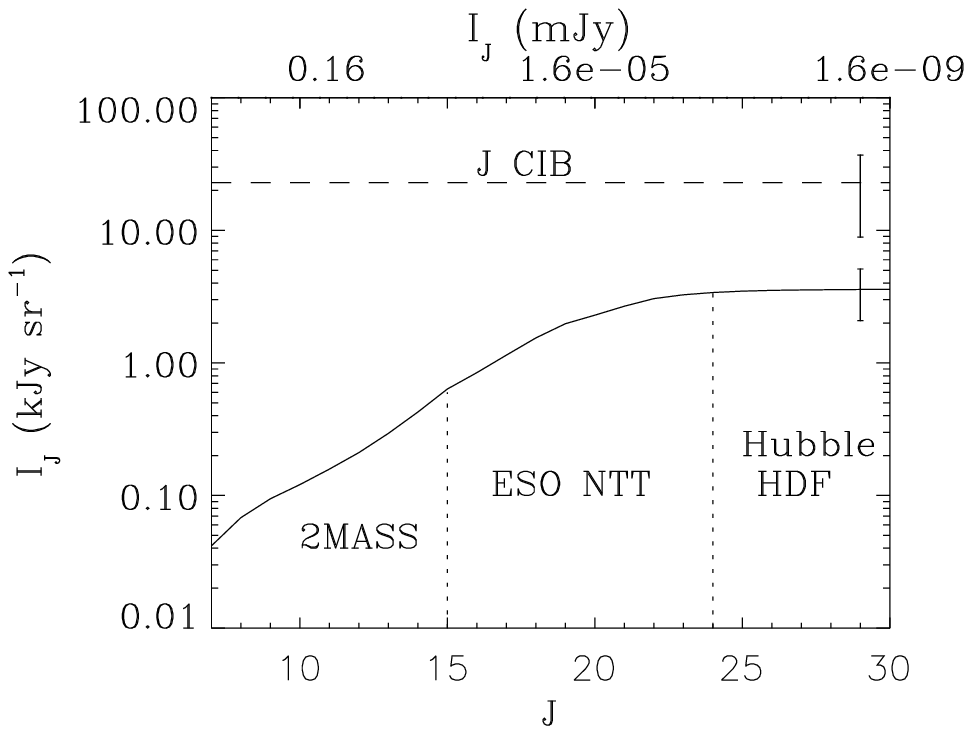}{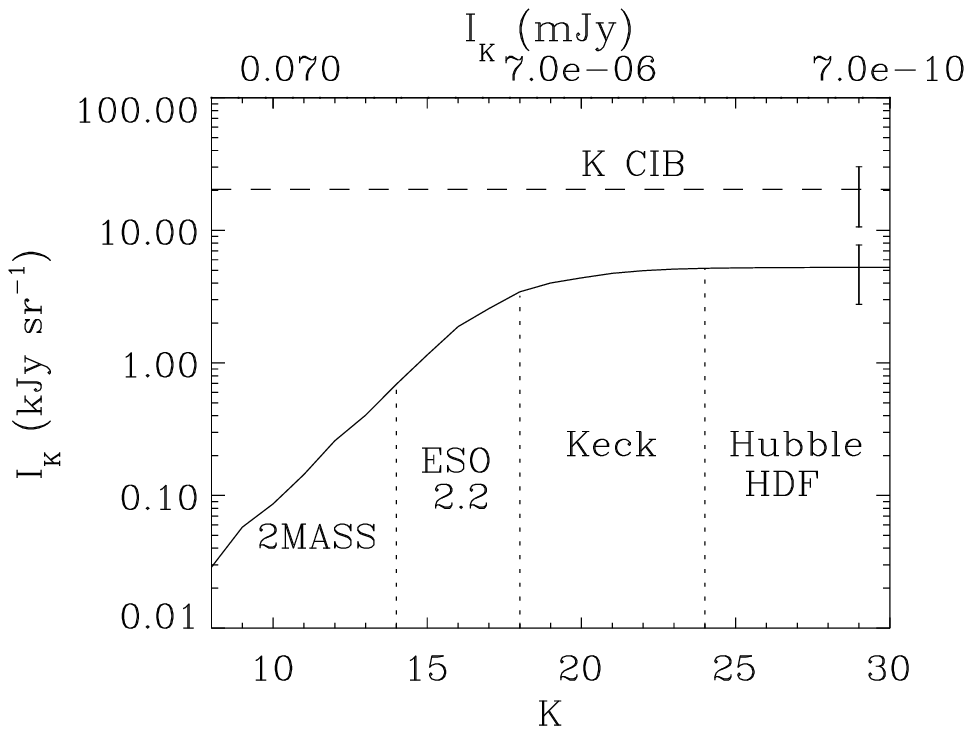}
\caption{Comparison of the CIB with the cumulative brightness derived from
	galaxy counts. Data are from 2MASS \citep{Jar01}, ESO NTT
	\citep{SDM+99}, ESO 2.2m \citep{SIG+97}, Keck \citep{BLK98} and HST
	\citep{TSW+99}. Error bars correspond to a 95\% ($2\sigma$)
	confidence level. A significant part of the energy is still not
	detected in sources for both colors.}
\label{galaxies}
\end{figure}

Figure~\ref{galaxies} shows the cumulative brightness of galaxies up to $J=30$
and $K=30$~mag. For the faintest $J$ and $K$ region (magnitudes 24 to 30) we
used the HST measurements assuming $m_{F160W}-m_J=0$ and $m_{F110W}-m_K = 0$.
The resulting error can be neglected, since the contribution of these very
faint galaxies to the total brightness is small (9\% of the total energy in
$J$, only 3\% in $K$). But note that 90\% of the energy is in the range
$12 \leq J \leq 24$ and $14 \leq K \leq 22$. 

Uncertainties on galaxy counts come from the Poisson statistics
($\sigma = \sqrt{N}$), and from completeness limitations and corrections. 
By comparing galaxy counts in the literature, we find that the number of
galaxies by range of magnitude $N$ is consistent to within a factor of 1.75,
which we adopt as a conservative value for the uncertainty.  
We find a $J$ total brightness for galaxies of $3.6\pm0.8$~kJy~sr$^{-1}$,
and a $K$ total brightness of $5.3\pm1.2$~kJy~sr$^{-1}$, in good agreement
with \citet{PMZ+98} who obtained 5.8~kJy~sr$^{-1}$ in the $K$ band.

The integrated galaxy counts are smaller than our DIRBE-2MASS CIB estimations,
even compared with the 95\% confidence level. 
We can address the possibility of large diffuse galaxy halos that would not be
included in the galaxy luminosities (because of their faintness) and we
find that to reduce the discrepancy between galaxy surface brightness and
our CIB value to less than $1\, \sigma$, an increase of each galaxy luminosity
by a factor 3.6 and 2.4 is required in $J$ and $K$, respectively.
If we assume that galaxy photometries are correct, a significant part of the
energy in the CIB is not detected in sources. The slope of the galaxy
luminosity function is observed to flatten for $K > 17$ \citep{SIG+97} and the
resulting slope for the cumulative brightness is almost 0 for $K>25$~mag. For
$J$, the slope change in the luminosity function is not as pronounced as for
$K$, but the slope of the cumulative brightness is also close to 0 for $J>25$.
Consequently, the slope of galaxy luminosity function must increase for
fainter magnitudes, or other contributions to the near-infrared background
must be invoked.  To summarize, the brightness of the cosmic background found
in this paper (Sect.~\ref{result}) is greater than the integrated brightness
of galaxy counts by
$19.3\pm7.8$~kJy~sr$^{-1}$ ($46.3\pm18.7$~nW~m$^{-2}$~sr$^{-1}$) at $J$ and
$15.1\pm6.1$~kJy~sr$^{-1}$ ($20.6\pm 8.3$~nW~m$^{-2}$~sr$^{-1}$) at $K$.

\subsection{Other possible contributions to the CIB}
Although galaxies appear to be unable to explain the brightness of the cosmic
background, there are other possible sources of energy production that could
lead to an isotropic cosmic background in the near-infrared.
An early burst of star formation inside or outside of galaxies, (that is,
either in primeval galaxies or in Population III stars), would result in a
nearly uniform and isotropic background \citep{BCH86}. The photospheric
emission from these stars would be redshifted into the infrared. 
For $z_*\sim 10 (100)$, the background due to high-mass stars will peak
at wavelengths 1 (10)~$\mu$m, and the background due to intermediate-mass
stars will be in the range 0.5--2 (6)~$\mu$m.
The gap between our background estimation and galaxy count contributions 
could be due to Population III stars. The present abundance of heavy
elements limits the amount of early star formation such that Population III
stars are not likely to be a very significant mass contribution to the
universe as a whole \citep{BCH86},
but the amount of star formation required to explain the near-infrared
cosmic background does not violate metallicity constraints. The implications
of such early star formation, for the small-scale density fluctuations of
primordial matter and for the nature of the intergalactic medium, would be
significant.

Depending on the power spectrum of small-scale density fluctuations, a
population of black holes could form before galaxies. The accretion
of surrounding material by these holes would generate energy that would make
an isotropic background today. 
The temperature of the accretion disk
would be $\sim 10^5$ K, and the accretion luminosity would be redshifted
into the near-infrared if the accretion is active at $z_{bh}\sim 30$--50
\citep{Car94}. 
Our observed $J$ band background brightness, after subtracting the
integrated light from galaxies, could be explained by the accretion
luminosity of a population of black holes.
Such black holes would not comprise enough dark matter to
close the universe, and their total mass would not violate 
constraints on the total baryon density inferred from models of 
primordial nucleosynthesis. But such a large
population of objects (comparable to the total mass of
galaxies) would contribute significantly to the total inventory
of baryons.

Massive particles that survive until redshift $z_\nu$ and subsequently
decay into photons will produce a background
of decay photons that is subsequently redshifted \citep{BCH86}.
For example, part of the $J$ and $K$ band background could be produced
by the decay of relic 100~eV neutrinos at redshift $z_\nu\sim 300$.
The number of plausible models is very large, and we can only 
speculate whether the background is due to such decaying particles.

\section{Conclusion}
\label{conclusion}
The integration of the 2MASS catalog over 550~deg$^{2}$ (i.e. $\sim 5500$
DIRBE pixels) leads to an estimate of the CIB for $J$ and $K$ of
$22.9\pm7.0$~kJy~sr$^{-1}$ ($54.0\pm16.8$~nW~m$^{-2}$~sr$^{-1}$) and
$20.4\pm4.9$~kJy~sr$^{-1}$ ($27.8\pm6.7$~nW~m$^{-2}$~sr$^{-1}$), respectively.

We have selected the most reliable areas of the sky by eliminating pixels
contaminated by bright stars, low galactic latitude regions, low ecliptic
latitude regions and possible cirrus clouds contribution.
A study of DIRBE$-$2MASS versus ecliptic latitude clearly shows that the
zodiacal light is responsible of most of the statistical uncertainty
($\sim 2$~kJy~sr$^{-1}$). The interplanetary dust model contains a
systematic error which is actually the dominant uncertainty in our results
($\sim 5$~kJy~sr$^{-1}$). 

The integrated brightness of known galaxies (down to the 30$^{\rm th}$
magnitude) represents only a fraction of the CIB.
Although the uncertainties due to residual contribution of the zodiacal light
are significant, the discrepancy between DIRBE$-$2MASS and galaxy counts would
suggest that a significant part of the energy is still not observed in discrete
sources. 
The slope of the galaxy luminosity functions may increase for galaxies fainter
than the 30$^{\rm th}$ magnitude, in which case the counts would agree with
the CIB values. However, models predict other possible contributions to the
background at these wavelengths \citep{BCH86} such as a burst of star formation
either in primeval galaxies or in Population III stars ($z\approx 10$),
very massive black holes (accreting from an uniform pre-galactic medium at
$z\approx 40$), massive decaying big bang relic particles ($z\approx 300$).
We have put new constraints in the near-infrared that encourage revisiting
the importance of those contributions to the CIB in cosmological models. 

The Space Infrared Telescope Facility (SIRTF) may extend this result for
the $L$ band at 3.5~$\mu$m and might be able to constrain the contributions of
the different components to the background.

The final all-sky 2MASS catalogs will permit a more accurate study of the
near-infrared excess: 100\% of the sky will be still available after
integration on the DIRBE beam, $\sim 75\%$ after bright star removal
(only 18\% for the 2MASS 2$^{\rm nd}$ Release presently used) and about
5000~deg$^{2}$ will remain at high galactic and ecliptic latitudes. 
This coverage should allow us to investigate structures in the zodiacal light
and may help to improve interplanetary dust models.

\acknowledgments
We are grateful to Roc Cutri for his help in accessing the 2MASS
data and for his critical reading of this paper and to George Helou for
helpful discussions.\\ 
This publication makes use of data products from the Two Micron All Sky Survey,
which is a joint project of the University of Massachusetts and the Infrared
Processing and Analysis Center/California Institute of Technology, funded by
the National Aeronautics and Space Administration and the National Science
Foundation.\\
\indent The COBE datasets were developed by the NASA Goddard Space Flight Center under
the guidance of the COBE Science Working Group and were provided by the
NSSDC.\\ 
\indent L. Cambr\'esy acknowledges partial support from the Lavoisier grant
of the French Ministry of Foreign Affairs.\\

\end{document}